\RequirePackage{fix-cm}
\documentclass[twocolumn,epjc3]{svjour3}  
\smartqed  

\RequirePackage{graphicx}
\usepackage[numbers]{natbib}
\usepackage{amssymb}
\usepackage{aas_macros}
\usepackage{changes}
\usepackage{hyperref}

\journalname{Eur. Phys. J. C}

\begin{document}

\title{Investigations of the Systematic Uncertainties in Convolutional Neural Network Based Analysis of Atmospheric Cherenkov Telescope Data}


\author{R.D. Parsons\thanksref{e1,addr1}
        \and
        A.M.W. Mitchell\thanksref{addr2,addr3} 
        \and
        S. Ohm \thanksref{addr4}
}

\thankstext{e1}{e-mail: daniel.parsons@physik.hu-berlin.de}


\institute{Institut f\"ur Physik, Humboldt-Universit\"at zu Berlin, Newtonstr. 15, D 12489, Berlin, Germany \label{addr1}
\and Department of Physics, ETH Zurich, CH-8093, Zurich, Switzerland \label{addr2}
\and Friedrich-Alexander-Universit\"at Erlangen-N\"urnberg, Erlangen Centre for Astroparticle Physics, Erwin-Rommel-Str. 1, D 91058 Erlangen, Germany \label{addr3}
\and DESY, Platanenallee 6, D 15738, Zeuthen, Germany\label{addr4}
}

\date{Received: date / Accepted: date}

\maketitle

\begin{abstract}

Machine learning, through the use of convolutional and recurrent neural networks is a promising avenue for the improvement of background rejection performance in imaging atmospheric Cherenkov telescopes. However, it is of paramount importance for science analysis that their performance remains stable against a wide range of observing conditions and instrument states.

We investigate the stability of convolutional recurrent networks by applying them to background rejection in a toy Monte Carlo simulation of a Cherenkov telescope array. We then vary a range of observation and instrument parameters in the simulation. In general, most of the resulting systematics are at a level not much greater than conventional analyses. However, a strong dependence of the neural network predictions on the noise level within the camera was found, with differences of up to 50\% in the gamma-ray acceptance rate in very noisy environments. It is clear from the performance differences seen in these studies that these observational effects must be considered in the training step of the final analysis when using such networks for background rejection in Cherenkov telescope observations.

\end{abstract}

\section{Introduction}
\label{sec:intro}

Rejection of the background of air showers triggered by charged cosmic rays is still the limiting factor in the majority of observations by imaging atmospheric Cherenkov telescopes (IACTs). Due to the comparatively much larger incidence rate of cosmic rays (a factor 10$^4$ even with the brightest sources), their identification and rejection constrains the sensitivity of Cherenkov telescopes. Identification of different primary particle species in Cherenkov telescopes relies on the observation of features in the Cherenkov camera image, which is essentially a projection of the shower development in the atmosphere onto a two-dimensional plane.

The most obvious differences in shower features, its width and length, have historically been extracted using \emph{Hillas Parameters} \cite{HillasParams}, taking the first and second moments of the camera images. However, even a cursory inspection of camera images reveals many details beyond this, with hadron induced air showers generally behaving in a more irregular way. Parameterisation of this irregularity can however be difficult due to the stochastic nature of air shower development.

Analysis of unparameterised camera images by advanced machine learning techniques such as convolutional neural networks is a promising avenue of research. In fact, a number of studies have shown that the background rate in both simulated arrays and real data \cite{RNNParsons,ShilonCNN} can be reduced by up to 25\%. 
Although such increases in performance are clearly an important step forward for IACT analysis, there are other factors which are not typically considered in the broader machine learning field which are important for their scientific usage, 
namely stability and reproducability of results. For Cherenkov telescopes this is most important in ensuring the stability of the classification performance (particularly the acceptance of gamma-ray events) under the diverse range of potential observing conditions. If classification methods are found to be unstable in performance, this would result not only in a mis-estimation of the sensitivity of the instrument, but an additional and potentially large source of systematic uncertainty in the measurement of photon flux from a source.

In this paper we construct an example image analysis scheme using convolutional recurrent neural networks, shown to perform well on H.E.S.S. data, and test its performance on a toy IACT model. We then introduce some of the most common variable factors
in observing and instrument conditions and measure the change in network performance, paying particular attention to the gamma-ray acceptance, i.e. the fraction of gamma-ray events passing the event selection cuts.

\section{Methodology and Simulations}
\label{sec:sims}

To get a fuller understanding of the effect of differing observation criteria on predictions of the background rejection method we take the approach of first defining a set of \emph{baseline} observing criteria. This is determined by selecting a telescope design, array layout and observing conditions, and using this as a comparison for all further tests. An investigation into the observing condition-dependent systematic uncertainties can then be made by the variation of individual parameters.

The traditional approach to such an investigation would be to create a full set of simulations using the combination of air shower simulations with a telescope ray-tracing and electronics simulation (such as the CORSIKA/\emph{sim\_telarray} combination \cite{simtel} used by CTA and H.E.S.S.). However, as the parameter space to be investigated increases, this approach requires a substantial investment of computing time in the simulation of the telescope response.

We therefore take the simpler approach of using the Cherenkov photon direction at ground level to produce a \emph{toy model} of a telescopes response \footnote{Toy simulation code available at \url{https://github.com/ParsonsRD/CORSIKA_toy_IACT}.}. By selecting all Cherenkov photons within a given telescope radius on the ground one can then simply bin their direction to produce the response of a perfect telescope at that position. More realism can then be achieved by the addition of efficiency factors to account for both the \emph{telescope mirror reflectivity} and \emph{photon detection efficiency} of the camera, reducing the weight of each simulated photon. Next, the photon directions are smoothed with the optical \emph{point spread function, PSF} of the telescopes and the photon weight is scattered randomly by the Gaussian \emph{photon detector resolution}. Finally a random Gaussian pedestal noise is  added around zero signal, to represent the noise from night sky background (assuming the contribution from electronic noise is negligible). The output of this toy telescope model is a series of realistic camera images representing detected photo-electrons (p.e.) which can then be used for reconstruction and classification.
\begin{table}
\begin{tabular}{lllll}
 & & & Post-cuts \\
Species & E$_{\rm min}$ & E$_{\rm max}$ & Showers\\

\hline
Gamma-ray & 30\,GeV & 300\,TeV &  $\sim120,000$ \\
Proton & 100\,GeV & 300\,TeV & $\sim120,000$\\

\end{tabular}
\caption{CORSIKA simulation energy ranges and statistics used in the data analysis.}
\label{tab:sims}
\end{table}


\begin{figure}[]
	\centering
	\includegraphics[width=\columnwidth]{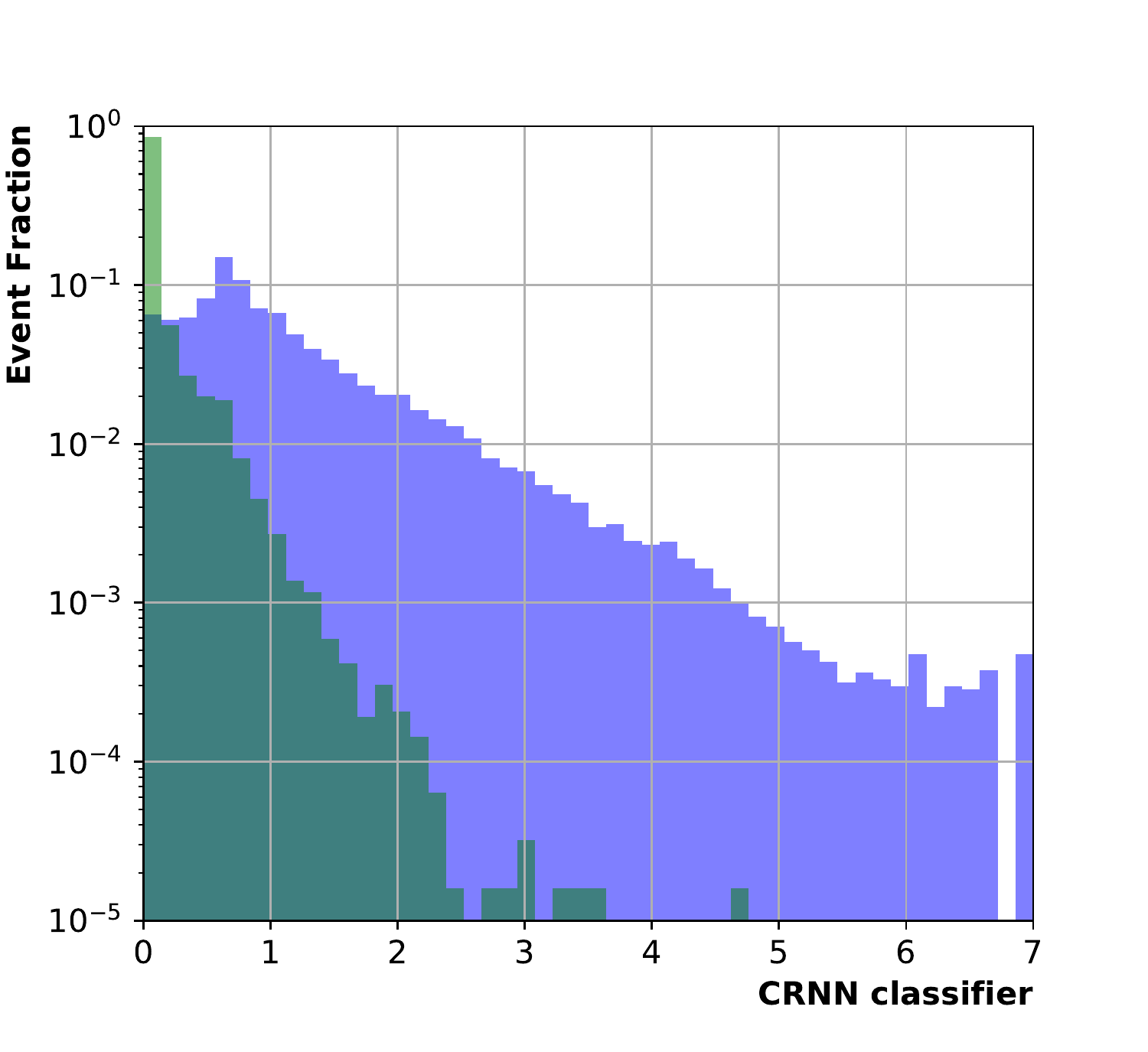}
	\caption{Distribution of the event classifier (calculated as $-$log$_{10}$[1-P$_\gamma$]) 
	for gamma-ray (weighted with a E$^{-2.6}$ spectrum) and proton events 
	(weighted with a E$^{-2.7}$ spectrum). }
	\label{fig:classifier}
\end{figure}

Taking this approach has the advantage that it significantly reduces the simulation time when investigating different telescope properties. Of course, the drawback here is that the predicted results are potentially less accurate and the size of systematic effects should be taken as qualitative only. However, as the quantitative size of systematic effects is likely strongly telescope and array design dependent, a study such as this would need to be repeated for any specific array design.

\subsection{Air Shower Simulations}
\label{sec:simulations}

To perform the training and testing of the neural networks, both, gamma-ray and proton air showers were simulated using CORSIKA \cite{CORSIKA} (version 76900) and the QGSJetII-04/UrQMD \cite{qgsjet,urqmd} hadronic interaction models. Events were simulated with an E$^{-2}$ spectral index and spread across a disc at ground level (1800\,m above sea level) of radius 6\,m. 
Cherenkov photons from particles in these showers were traced to ground level and subjected to atmospheric extinction, then passed to the toy telescope simulation. Details for the simulation of both particle species are given in table \ref{tab:sims}. Different lower bounds for the energy of the simulations were chosen for gamma-ray and proton simulations, due to the differing fractions of energy put into the production of electrons in the two shower types, with proton showers typically producing around a third of the Cherenkov yield at ground level.

\subsection{Baseline Array}
\label{sec:baseline}

\label{sec:network}

For the baseline array design, a 3$\times$3 grid of telescopes with a spacing of 120\,m was chosen. The individual telescopes created have a 12\,m diameter, a square camera with an 8$^\circ\times$8$^\circ$ field of view and a pixel width of 0.2$^\circ$ -- similar to 
the CTA medium size telescopes \cite{MST}. 
Telescopes were simulated with a 80\% mirror reflectivity and a 20\% photon detection efficiency of the photomultiplier tubes.
Additionally, a uniform Gaussian optical PSF of 0.04$^\circ$ was applied to the images, which is typical for the centre of the field of view of a Davis-Cotton optical configuration \cite{DC}. Finally a randomly assigned pedestal value chosen from a Gaussian of width one p.e. was added to the images.

The sections of the images without a Cherenkov signal were then cleaned away using a split level tailcut algorithm \cite{HESSCrab} and the Hillas moments of the images calculated \cite{HillasParams}. Images with a summed amplitude in the cleaned images of less than 80 p.e. or with its centroid position more than 3$^\circ$ from the camera centre were discarded. The remaining images were then used to reconstruct the direction and core position of the event using the intersection of the Hillas parameter major axes (described in \cite{HESSCrab}) using routines from the \emph{ctapipe} software library \cite{ctapipe}. 

\subsection{Recurrent Network Design}

The convolutional recurrent neural network  structure used for background rejection is essentially a reimplemention of the proven network design presented in \cite{RNNParsons} using the \emph{Keras}\,\cite{Keras} machine learning framework, with the \emph{TensorFlow} backend\,\cite{Tensorflow}. This network contains two inputs, the direct camera images (re-normalised such that their peak amplitude is equal to 1), and the  parameterised input, namely the image width, length, amplitude, displacement of centroid from source position, offset of the centroid in the field of view and reconstructed impact parameter\footnote{Network training and evaluation code available at \url{https://github.com/ParsonsRD/CRNN_trainer}}.

Image inputs were then processed by the typical alternation of convolutional and pooling layers and then flattened and fed through a recurrent layer (specifically a long short-term memory (LSTM) layer \cite{LSTM}) with each telescope image acting as a ``time step'' of the input.
Parameterised inputs were  similarly processed by a recurrent layer then the two input streams are concatenated, again processed through a recurrent layer and passed to the output layer. At several points within the network dropout layers (with a 50\% dropout rate) are included to regularise the training performance. 
The network was then trained using 50\% of the simulated data as input using the \emph{categorical cross-entropy} loss function. To simplify the interpretation of results the training was performed across the full energy range, rather than splitting into smaller energy bins as previously investigated. The training was terminated if the network goes 50 training epochs without improvement in performance (evaluated on a subset of the training data) reverting to the training epoch with the best performance. Generally, training was completed after around 300 training epochs were performed.

Once training was complete the remaining data sample was then used to assess the network performance.

\section{Investigation of Systematic Effects}
\label{sec:systematics}

\subsection{Night Sky Background Light}

\begin{figure*}[h]
	\centering
	\includegraphics[width=\columnwidth]{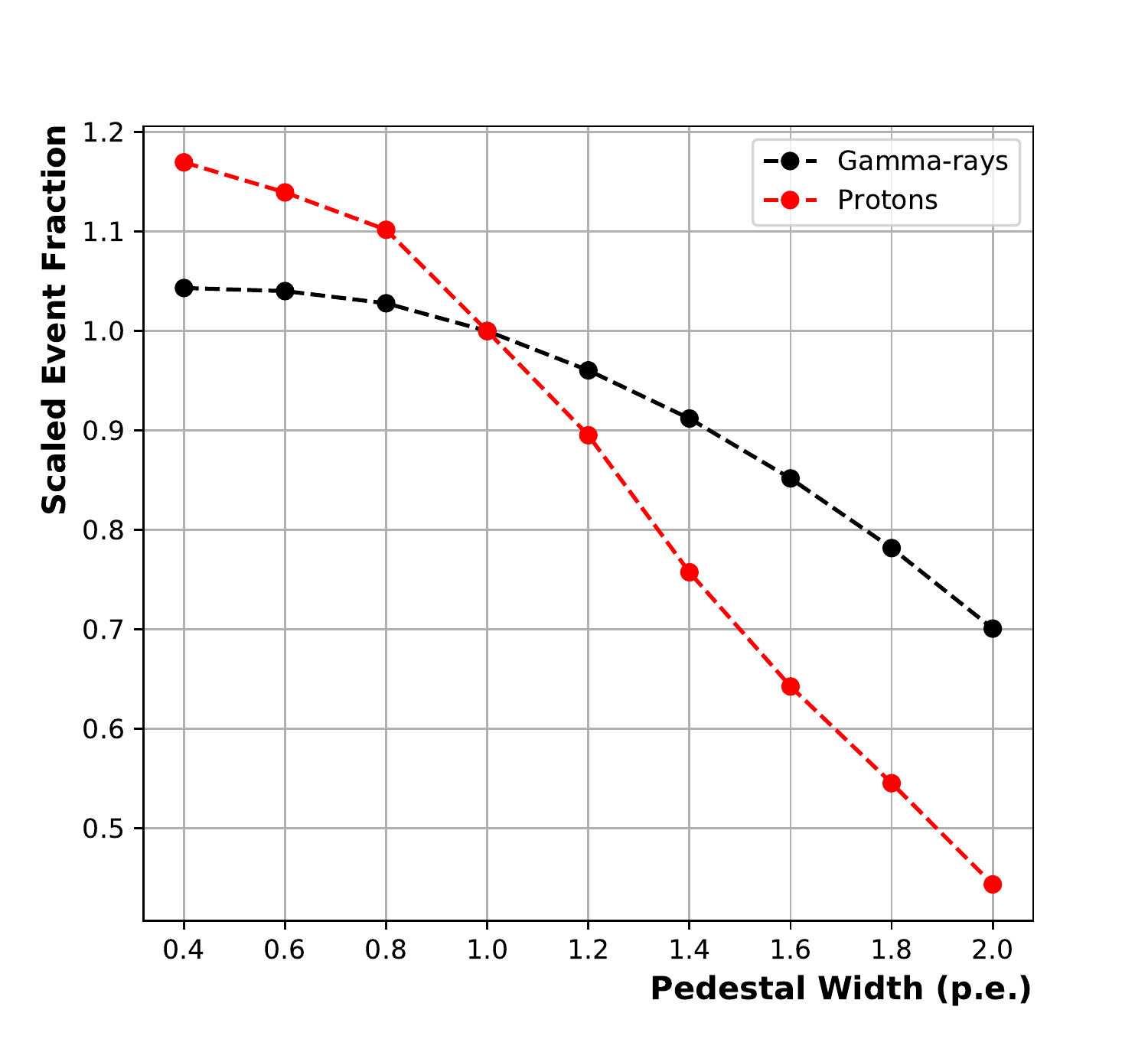}
	\includegraphics[width=\columnwidth]{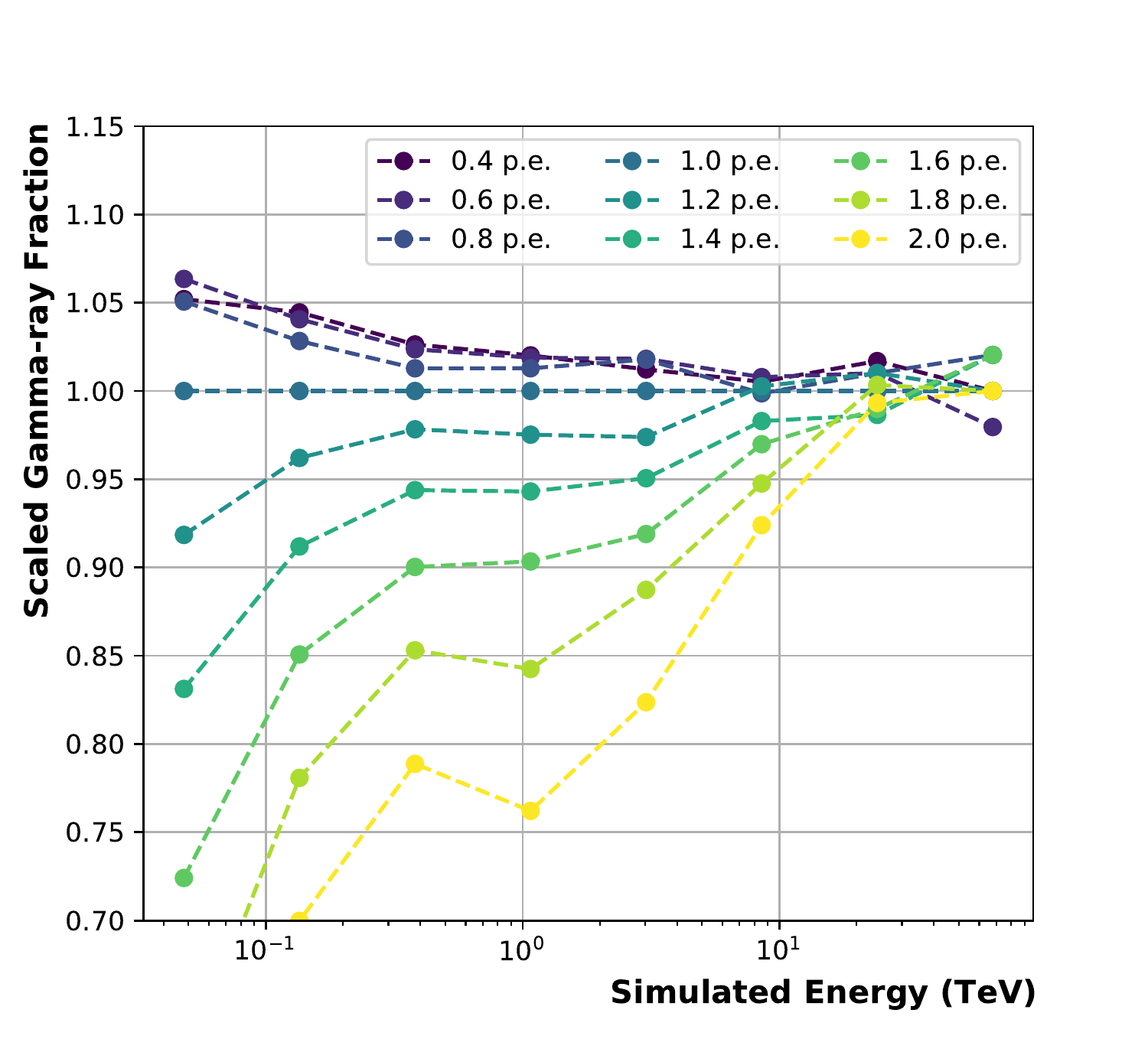}
	\caption{Acceptance change of the CRNN classifier with differing noise levels 	(pedestal width in number of p.e.). Left: Shown for both gamma-ray and proton events integrated over all energies (weighted with a E$^{-2.6}$ and  E$^{-2.7}$ spectrum respectively). 	Right: Shown for gamma-rays as a function of energy, differing noise levels are represented by the colour scale. }
	\label{fig:NSBperformance}
\end{figure*}

As shown in previous investigations \cite{RNNParsons} the level of noise in the image is clearly important to the performance of the network. Figure \ref{fig:NSBperformance} shows the gamma-ray acceptance of the recurrent network as a function of the pedestal width (proportional to the square root of NSB noise level). In comparison with the previously studied networks a similar reduction in acceptance is seen as the NSB level increases. The rejection of background events increases similarly to  the  fall in gamma-ray acceptance, showing that generally the additional random noise in the images pushes the network to classify them as background-like.

Figure \ref{fig:NSBperformance} demonstrates that this effect is most pronounced at low energies. Tests with lower than nominal NSB, show a slight increase in acceptance compared to the nominal noise level and once energies of around 500\,GeV are reached performance is identical. At higher NSB levels the energy dependence is quite pronounced with very large differences in acceptance seen at 100\,GeV. As energy increases the behaviour of the different NSB levels gradually converge, with a roughly consistent acceptance being seen at around 10\,TeV. This evolution with energy is expected simply due to the generally increased brightness of the camera images as the energy, with the level of noise becoming smaller relative to the brightness of the shower images. It is therefore unsurprising that even the highest noise levels have little effect on the performance above 10\,TeV.


\subsection{Zenith Angle}

\begin{figure*}[t]
	\centering
	
	\includegraphics[width=\columnwidth]{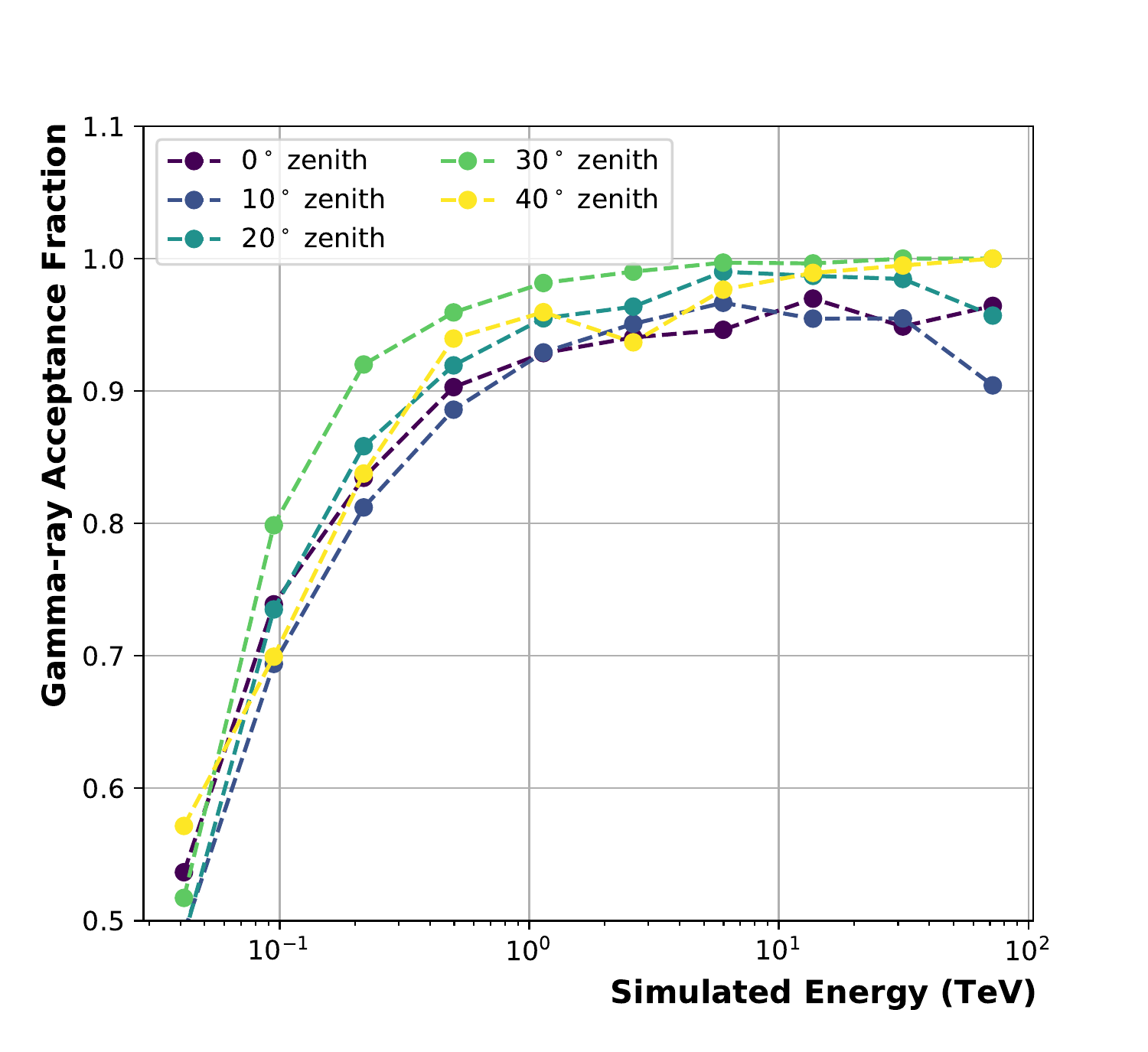}
	\includegraphics[width=\columnwidth]{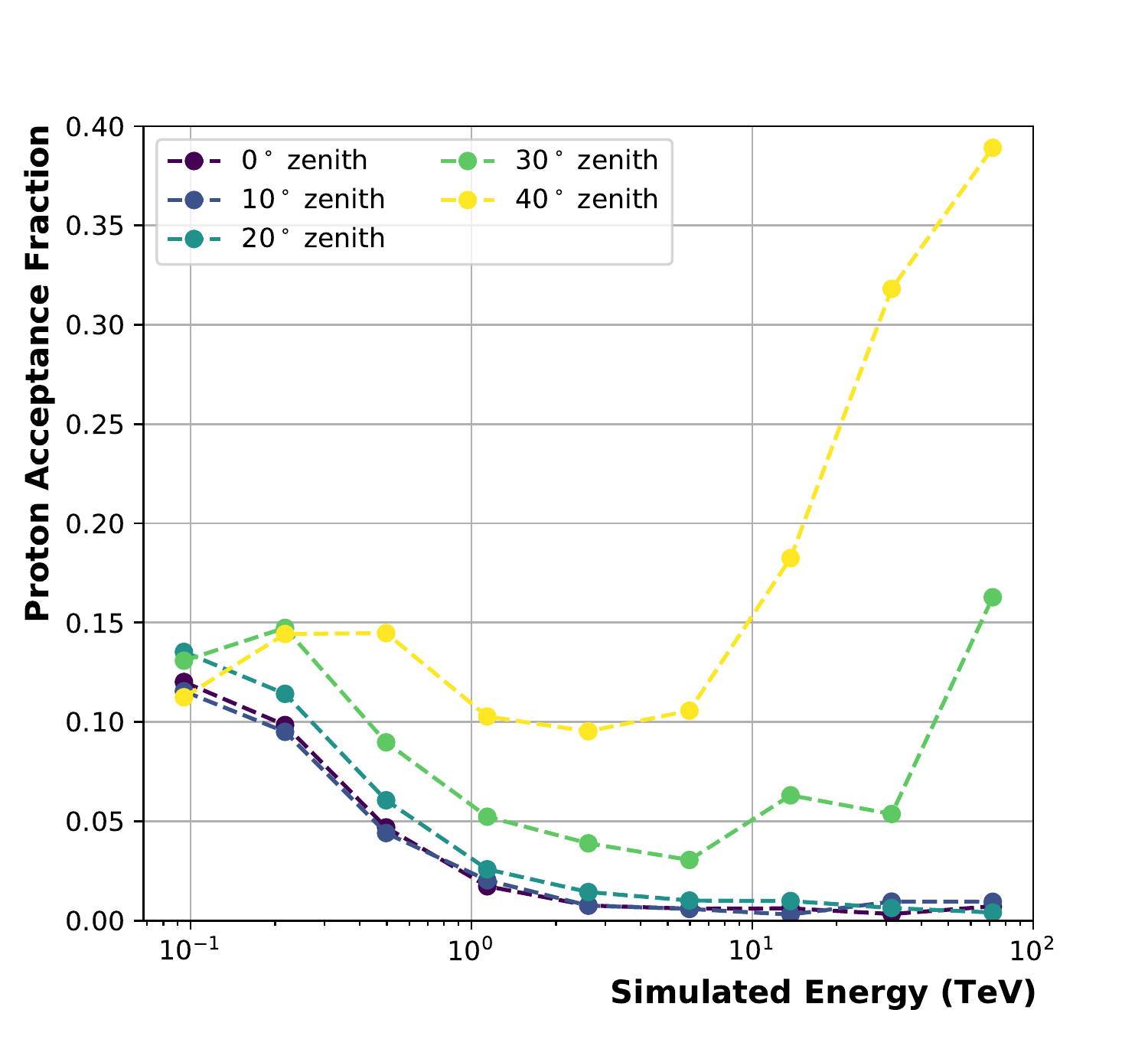}

	\caption{Energy dependant acceptance, defined as the fraction of preselected events passing background rejection cuts, for both gamma-rays (left) and protons (right) at several zenith angles using a classifier trained with vertical events.}
	\label{fig:Zenithperformance}
\end{figure*}

When assessing the stability of the analysis it is also important to consider that the background rejection (at least in the current generation of instruments) is typically trained in fixed zenith angle bands and then its outcome interpolated between these bands. It is therefore important to understand the sensitivity of the classification prediction with zenith angle to ensure a smooth transition in performance between zenith bands.

To test the network stability, gamma-ray and proton simulations were run from 0$^\circ$-40$^\circ$ zenith angle. The larger zenith angle simulation results were then classified using the neural network trained using vertical showers, and a cut value defined using the vertical showers (80\% acceptance). Figure \ref{fig:Zenithperformance} shows the energy dependent gamma-ray and proton acceptances for the different zenith angles. This figure shows the gamma-ray acceptance to be remarkably stable across all zenith angles, with variations of only 5-10\% seen. However, this lack of variation is likely due to the relatively high acceptance of the cuts used, this can be seen clearly in the proton acceptance where a significant reduction in rejection performance (around a factor 5) is seen at 30$^\circ$ zenith angle. This implies that at larger zenith angles the gamma-ray acceptance stability is achieved by classifying many more events as gamma-rays.

Only minor degradation in performance is seen at 20$^\circ$ zenith angle however, which if one assumes a fixed spacing of training bands in steps of $\Delta \sec \theta_{\rm zen} = 0.06$, it would require around 32 training bands to cover the typical range of observations up to 70$^\circ$ zenith.



\subsection{Deactivated Pixels}

\begin{figure}[t]
	\centering
	\includegraphics[width=\columnwidth]{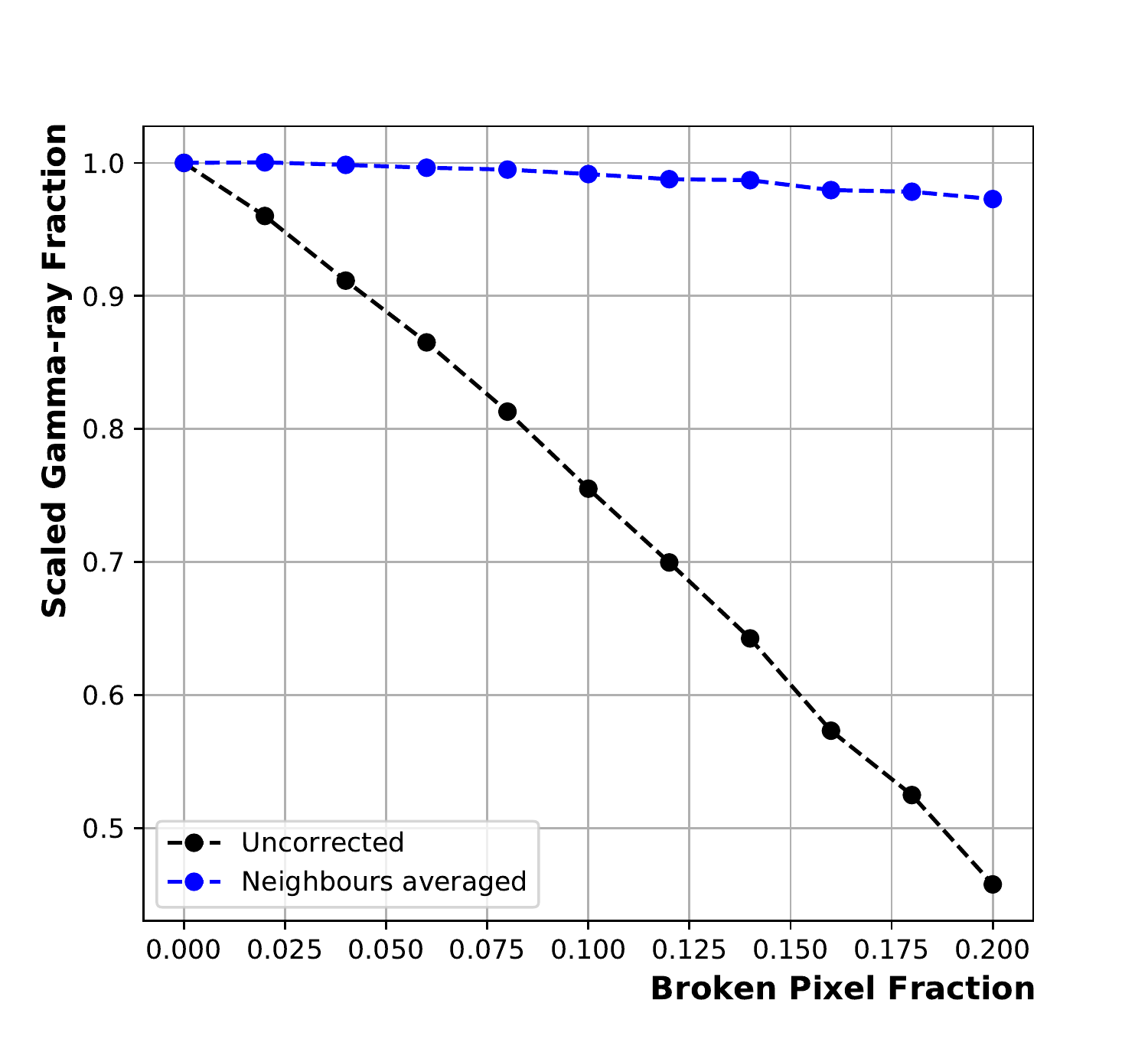}
	\caption{Acceptance change of gamma-ray events as a function of the fraction of deactivated camera pixels both with and without the averaging of neighbouring pixels (weighted with a E$^{-2.6}$ spectrum).}
	\label{fig:BPperformance}
\end{figure}

When operating Cherenkov telescopes it is not unusual for a fraction of the pixels in the camera to return no signal in any given observation run. These deactivations could be caused either by problems with the camera hardware or bright stars in the field of view tripping safety measures. In most observations only a small number of camera pixels are affected and hence deactivated pixels are typically ignored when calculating Hillas parameters. However, in this case where we are analysing the full camera pixel-wise image information there is no easy way of ignoring a given camera pixel. 

To simulate the deactivation of pixels in an observation, a fraction of image pixels fed to the convolutional layers are randomly chosen and set to an intensity of zero. Figure \ref{fig:BPperformance} shows the reduction of gamma-ray acceptance as a function of broken pixel fraction. Introducing these broken pixels has a strong effect on the network performance with only a 5\% broken pixel fraction reducing the gamma-ray acceptance by around 20\%. However, if one performs a more sensible infill of these inactive pixels by simply allocating them with the average of the four surrounding pixels, the effect of deactivated pixels is significantly reduced, producing a modification of the gamma-ray acceptance of only the few percent level. 

\subsection{Optical Point Spread Function}

\begin{figure}[]
	\centering
	\includegraphics[width=\columnwidth]{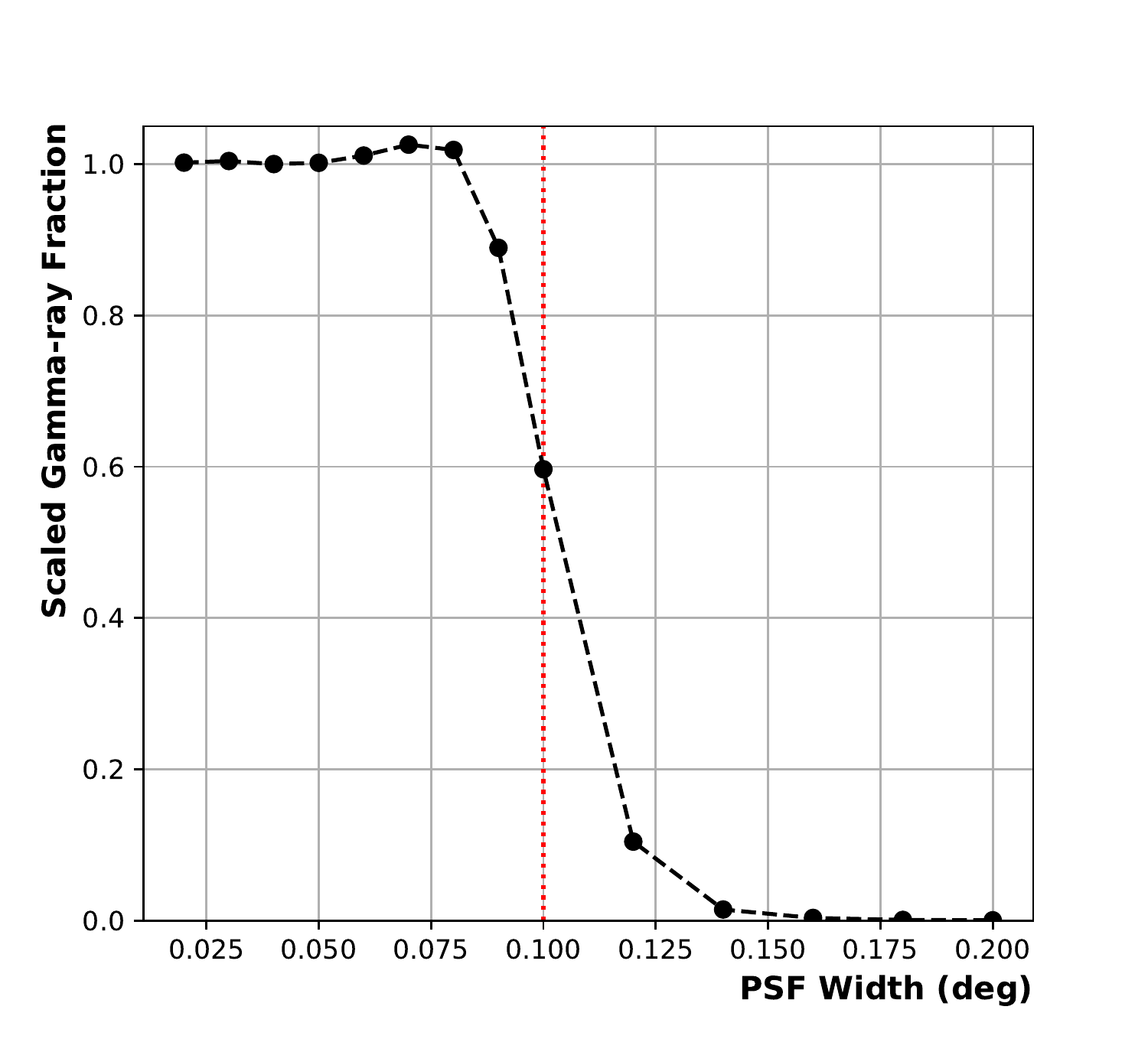}
	\caption{Acceptance change of gamma-ray events as a function of the width of the optical PSF (weighted with a E$^{-2.6}$ spectrum). The red vertical line shows the pixel radius simulated.
	   }	
	   \label{fig:PSFperformance}
\end{figure}

Variations of the optical PSF were tested from 0.01$^\circ$ to 0.2$^\circ$. Figure \ref{fig:PSFperformance} shows that at low PSF values the effect of changing the PSF is negligible. However, as the PSF size approaches the pixel size (0.1$^\circ$ radius) the acceptance of the baseline neural network falls off rapidly. This behaviour can be understood as at small values of optical PSF the pixelisation is the limiting factor of the image resolution, so changes in PSF value have little effect on the morphology. Once the PSF becomes similar to the pixel size however, the image morphology begins to change, significantly affecting network performance.

It should be noted that in a real telescope the optical PSF is not constant across the field of view as in our toy model. Instead the PSF is typically at its best at the centre of the field of view, gradually worsening with offset angle. For example the 80\% containment radius of photons in the smaller H.E.S.S. telescopes reaches the pixel size at around 2$^\circ$ from the optical axis \cite{HESSoptics}. Therefore it is quite possible that the images from sources simulated at different offsets perform differently in the network, which may require additional training sets to handle well or a careful selection of the training data used.

\section{Discussion}

The study of changing observing conditions on the behaviour of convolutional neural networks used for background rejections in IACTs has shown several interesting effects on the network performance. Firstly, the most significant effect seems to originate from the changing level in NSB, most significantly at larger noise levels. Clearly this dependence on NSB level represents a challenge to the analysis of Cherenkov telescope data, where noise levels can change dramatically from one observing position to another and even within the field of view when looking at bright sky regions such as the Galactic Centre. It may be possible to mitigate some of the negative effect of changing noise levels by using de-noising algorithms (for example wavelet de-noising or de-noising autoencoders) to reduce impact of the noise on data. However, ultimately it may be required to directly simulate the specific set of conditions applicable to observations when analysing some datasets, although at high energies where the noise level is not so important this would not be required.

Deactivating pixel outputs also has a strong effect on performance (although with next generation instruments any more than a few percent of pixels being inactive would be very unusual), however their effect is easily mitigated to the few percent level by a simple averaging of the neighbouring pixel contents. Potentially, this effect could be further reduced by a more intelligent reconstruction of the surrounding pixels.

To keep the systematic uncertainties introduced by using a CRNN trained to mis-matched conditions at the less than $\sim10\%$ level, this study shows i) that the noise must be accurate to within $\sim 0.5$p.e., ii) that the zenith angle must be accurate to within a step of 0.06 in $\sec\theta_{\rm zen}$, iii) that deactivated pixels must be in-filled prior to event reconstruction, and iv) that the optical PSF must be as accurately modelled as possible once it extends beyond the size of a single pixel. 


Several more detailed effects that may adversely affect the network performance (yet at a level likely less dramatic as those explored here) include the variation of performance with off-axis gamma-ray source; an asymmetric PSF shape that varies over the field of view; degradation of the optical throughput of the telescopes and variation in atmospheric conditions. These may be explored at a later stage, once the uncertainties due to the primary variables can be considered sufficiently under control. Additionally, similar studies to that presented here should be performed for each unique atmospheric Cherenkov telescope design and array layout under consideration, in order to determine the necessary accuracy of the simulations for optimal analysis performance.

Ultimately, the decision to employ CRNNs as part of the analysis of IACT data will be a trade-off between achieving the best possible performance and the computing effort and time needed to run appropriate simulations and train the network whilst keeping systematic uncertainties under the required levels.

\section*{Acknowledgements}
\noindent AM is supported by the Deutsche Forschungsgemeinschaft (DFG, German Research Foundation) – Project Number 452934793, MI 2787/1-1.

\bibliographystyle{number_cite}

\bibliography{RNN}

\end{document}